\documentclass[showpacs,10pt,twocolumn,prb]{revtex4-1}

\usepackage{amsmath}
\usepackage{amssymb}
\usepackage{graphicx}
\usepackage{amssymb}
\usepackage{graphics}
\usepackage{epsfig}
\usepackage{CJK}
\usepackage{color}

\begin{document}

\begin{CJK*}{GBK}{Song}
\title{Anisotropic magnetocaloric effect and critical behavior in CrSbSe$_3$}
\author{Yu Liu,$^{1}$ Zhixiang Hu,$^{1,2}$ and C. Petrovic$^{1,2}$}
\affiliation{$^{1}$Condensed Matter Physics and Materials Science Department, Brookhaven National Laboratory, Upton, New York 11973, USA\\
$^{2}$Materials Science and Chemical Engineering Department, Stony Brook University, Stony Brook, New York 11790, USA}
\date{\today}

\begin{abstract}
We report anisotropic magnetocaloric effect and critical behavior in quasi-one-dimensional ferromagnetic CrSbSe$_3$ single crystal. The maximum magnetic entropy change $-\Delta S_M^{max}$ is 2.16 J kg$^{-1}$ K$^{-1}$ for easy $a$ axis (2.03 J kg$^{-1}$ K$^{-1}$ for hard $b$ axis) and the relative cooling power $RCP$ is 163.1 J kg$^{-1}$ for easy $a$ axis (142.1 J kg$^{-1}$ for hard $b$ axis) near $T_c$ with a magnetic field change of 50 kOe. The magnetocrystalline anisotropy constant $K_u$ is estimated to be 148.5 kJ m$^{-3}$ at 10 K, decreasing to 39.4 kJ m$^{-3}$ at 70 K. The rescaled $\Delta S_M(T,H)$ curves along all three axes collapse onto a universal curve, respectively, confirming the second order ferromagnetic transition. Further critical behavior analysis around $T_c \sim$ 70 K gives that the critical exponents $\beta$ = 0.26(1), $\gamma$ = 1.32(2), and $\delta$ = 6.17(9) for $H\parallel a$, while $\beta$ = 0.28(2), $\gamma$ = 1.02(1), and $\delta$ = 4.14(16) for $H\parallel b$. The determined critical exponents suggest that the anisotropic magnetic coupling in CrSbSe$_3$ is strongly dependent on orientations of the applied magnetic field.
\end{abstract}
\maketitle
\end{CJK*}

\section{INTRODUCTION}

Low-dimensional ferromagnetic (FM) semiconductors, holding both ferromagnetism and semiconducting character, form the basis for nano-spintronics application. Recently, the two-dimensional (2D) CrI$_3$ and Cr$_2$Ge$_2$Te$_6$ have attracted much attention since the discovery of intrinsic 2D magnetism in mono- and few-layer devices.\cite{Huang,McGuire0,McGuire,Gong} Intrinsic magnetic order is not allowed at finite temperature in low-dimensional isotropic Heisenberg model by the Mermin-Wagner theorem,\cite{Mermin} however, a large magnetocrystalline anisotropy removes this restriction, for instance, the presence of a magnetically ordered state in the 2D Ising model. The enhanced fluctuations in 2D limit make symmetry-breaking order unsustainable, however by gapping the low-energy modes through the introduction of anisotropy, order could be established by providing stabilization of long-range correlations in 2D limit. Given the reduced crystal symmetry in low-dimensional magnets, an intrinsic magnetocrystalline anisotropy can be expected and points to possible long-range magnetic order in atomic-thin limit.

In ternary chromium trichalcogenides, Cr(Sb,Ga)X$_3$ (X = S, Se, Te) displays a pseudo-one-dimensional crystal structure. This is different from Cr(Si,Ge)Te$_3$ that features layered structure and a van der Waals bonds along the $c$-axis. In Cr(Sb,Ga)X$_3$, the CrX$_6$ octahedra form infinite, edge-sharing, double rutile chains. The neighboring chains are linked by Sb or Ga atoms. The FM semiconductor CrSbSe$_3$ has attracted considerable attention.\cite{Volkov,Odink,Kong,Yan} A band gap of 0.7 eV was determined by resistivity and optical measurements.\cite{Kong} The Cr in CrSbSe$_3$ exhibits a high spin state with $S$ = 3/2, and orders ferromagnetically below the Curie temperature $T_c$ of 71 K.\cite{Kong} FM state in CrSbSe$_3$ is fairly anisotropic with the $a$ axis being the easy axis and the $b$ axis being the hard axis. The critical analysis where magnetic field was applied along the $a$ axis suggests that the ferromagnetism in CrSbSe$_3$ cannot be simply described by the mean-field theory.\cite{LIUYU,Gao} This invites the detailed investigation on its anisotropic critical behavior.

The magnetocaloric effect (MCE) can give additional insight into the magnetic properties, and it could be also used to assess magnetic refrigeration potential.\cite{YuLIU,Xiao,Liu01,YanJ,Fu,Sun,Liu02,Liu03,Liu04} Bulk CrSiTe$_3$ exhibits anisotropic entropy change ($-\Delta S_M^{max}$) with the values of 5.05 and 4.9 J kg$^{-1}$ K$^{-1}$ at 50 kOe for out-of-plane and in-plane fields, respectively, with the magnetocrystalline anisotropy constant $K_u$ of 65 kJ m$^{-3}$ at 5 K.\cite{Liu01} The values of $-\Delta S_M^{max}$ are about 4.24 J kg$^{-1}$ K$^{-1}$ (out-of-plane) and 2.68 J kg$^{-1}$ K$^{-1}$ (in-plane) at 50 kOe for CrI$_3$ with a much larger $K_u$ of 300 $\pm$ 50 kJ m$^{-3}$ at 5 K.\cite{Richter} The large magnetocrystalline anisotropy is important in preserving FM in the 2D limit.

In the present work we investigate the anisotropic magnetic properties of pseudo-one-dimensional CrSbSe$_3$ single crystals. The magnetocrystalline anisotropy constant $K_u$ is strongly temperature-dependent.It takes a value of $\sim$ 148.5 kJ m$^{-3}$ at 10 K and monotonically decreases to 39.4 kJ m$^{-3}$ at 70 K for the hard $b$ axis. The $K_u$ of CrSbSe$_3$ is much larger than that of Cr$_2$(Si,Ge)$_2$Te$_6$ but comparable with that of Cr(Br,I)$_3$. This results in anisotropic magnetic entropy change $\Delta S_M(T,H)$ and relative cooling power (RCP), as well as in magnetic critical exponents $\beta$, $\gamma$, and $\delta$ that point to the nature of the phase transition. The anisotropic magnetic coupling of CrSbSe$_3$ is strongly dependent on orientations of the applied magnetic field, providing an excellent platform for further theoretical studies of low-dimensional magnetism.

\section{EXPERIMENTAL DETAILS}

CrSbSe$_3$ single crystals were fabricated by the self-flux technique starting from an intimate mixture of raw materials Cr (99.95\%, Alfa Aesar) powder, Sb (99.999\%, Alfa Aesar) pieces, and Se (99.999\%, Alfa Aesar) pieces with a molar ratio of 7 : 33 : 60. The starting materials were sealed in an evacuated quartz tube and then heated to 800 $^\circ$C and slowly cooled to 680 $^\circ$C with a rate of 2 $^\circ$C/h. Needle-like single crystals with lateral dimensions up to several millimeters can be obtained. The element analysis was performed using an energy-dispersive x-ray spectroscopy in a JEOL LSM-6500 scanning electron microscope (SEM), confirming a stoichiometric CrSbSe$_3$ single crystal. The powder x-ray diffraction (XRD) data were taken with Cu K$_{\alpha}$ ($\lambda=0.15418$ nm) radiation of Rigaku Miniflex powder diffractometer. The anisotropy of magnetic properties were measured by using one single crystal with mass of 0.32 mg and characterized by the magnetic property measurement system (MPMS-XL5, Quantum Design). The applied field ($H_a$) was corrected as $H = H_a-NM$, where $M$ is the measured magnetization and $N$ is the demagnetization factor. The corrected $H$ was used for the analysis of magnetic entropy change and critical behavior.

\section{RESULTS AND DISCUSSIONS}

\begin{figure}
\centerline{\includegraphics[scale=1]{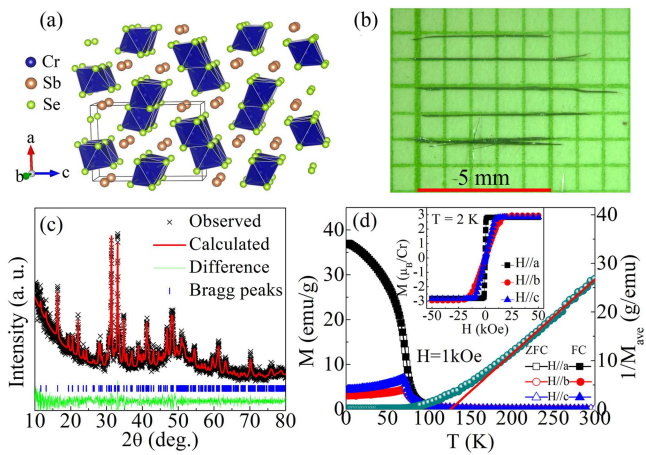}}
\caption{(Color online). (a) Crystal structure of CrSbSe$_3$ and (b) representative single crystals on a millimeter-grid paper. (c) Powder x-ray diffraction (XRD) pattern of CrSbSe$_3$. (d) Temperature-dependent magnetization M(T) measured in H = 1 kOe with zero field cooling (ZFC) and field cooling (FC) modes along all three axes (left axis) and inverse average magnetization $1/M_{ave} = 3/(M_a + M_b +M_c)$ (right axis) fitted by the Curie-Weiss law. Inset shows the field-dependent magnetization $M(H)$ at 2 K.}
\label{1}
\end{figure}

Figure 1(a) displays the CrSbSe$_3$ crystal structure. The material crystalizes in an orthorhombic lattice with the space group of $Pnma$. That is a pseudo-one-dimensional structure with double rutile chains of CrSe$_6$ octahedra that are aligned parallel to the $b$ axis. As shown in Fig. 1(b), the $b$ axis is along the long crystal dimension in single crystals. Within the double chain, the Cr cations form an edge-sharing triangular arrangement, while the Sb atoms link the adjacent chains. In the powder XRD pattern [Fig. 1(c)], all peaks can be well indexed by the orthorhombic structure (space group $Pnma$) with lattice parameters $a$ = 9.121(2) {\AA}, $b$ = 3.785(2) {\AA} and $c$ = 13.383(2) {\AA}, in good agreement with previous report.\cite{Kong}

Figure 1(d) shows the temperature dependence of magnetization $M(T)$ along all three axes measured at $H$ = 1 kOe. There is no bifurcation seen between the zero-field-cooling (ZFC) and field-cooling (FC) curves, indicating the high quality of single crystal. The $M(T)$ curves are nearly isotropic at high temperature but show an obvious anisotropic magnetic response for field applied along different axes at low temperature. For $H\parallel a$, a rapid increase near 70 K in $M(T)$ on cooling corresponds well to the reported paramagnetic (PM) to FM transition.\cite{Kong} For $H\parallel b$ and $H\parallel c$, an anomalous peak feature is observed, which is also seen in Cr$_2$(Si,Ge)$_2$Te$_6$ and Cr(Br,I)$_3$ with a large magnetocrystalline anisotropy.\cite{Liu05,Casto,Selter} The inverse average magnetization $1/M_{ave} = 3/(M_a+M_b+M_c)$ can be well fitted from 200 to 300 K by using the Curie-Weiss law, which generates an effective moment of 4.6(1) $\mu_B$/Cr and a positive Weiss temperature of 125(1) K, in line with the values reported for CrSbSe$_3$ polycrystals.\cite{Volkov,Odink} The anisotropic magnetization isotherms measured at $T$ = 2 K [inset in Fig. 1(d)] show a similar saturated magnetization $M_s$ at 3 $\mu_B$/Cr, consistent with expectation of $S = 3/2$ for Cr$^{3+}$, but different saturated fields $H_s$ of 1 kOe, 18 kOe, and 12 kOe for $a$, $b$, and $c$ axis, respectively, close to the values in the previous reports.\cite{Kong,Yan}

\begin{figure}
\centerline{\includegraphics[scale=1]{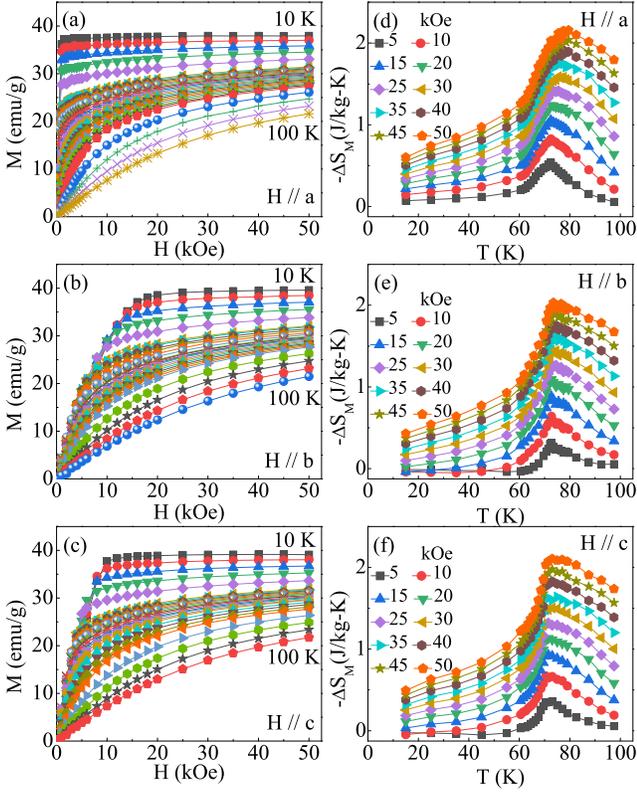}}
\caption{(Color online). (a-c) Typical initial isothermal magnetization curves measured along all three axes with temperature ranging from 10 to 100 K. (d-f) The corresponding calculated magnetic entropy change $-\Delta S_M(T)$ at various fields change.}
\label{2}
\end{figure}

\begin{figure}
\centerline{\includegraphics[scale=1]{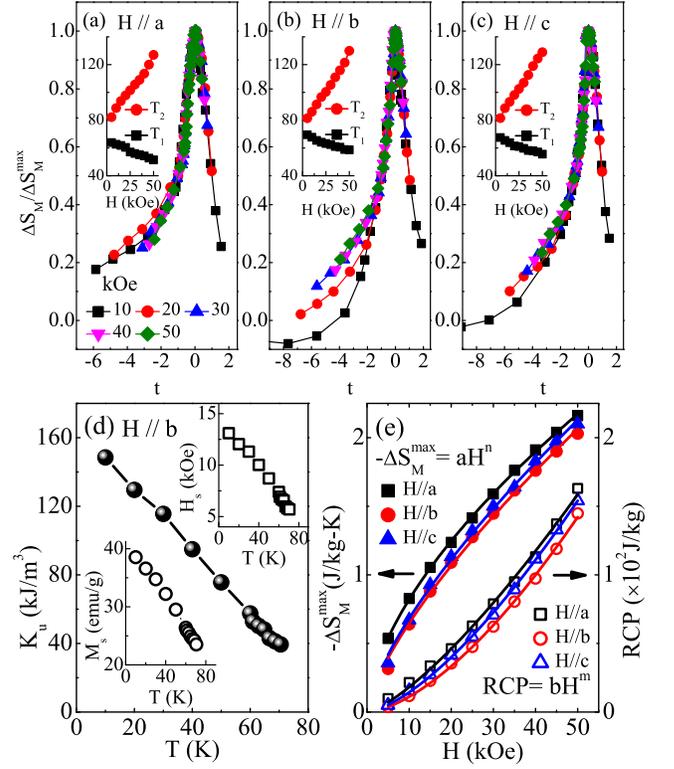}}
\caption{(Color online). (a-c) Normalized magnetic entropy change $\Delta S_M$ as a function of the reduced temperature $t$ along all three principal crystallographic axes of CrSbSe$_3$. Insets show the evolution of the reference temperatures $T_1$ and $T_2$. (d) Temperature dependence of the magnetocrystalline anisotropy constant $K_u$, the saturation field $H_s$, and the saturation magnetization $M_s$ (insets) estimated from the hard $b$ axis below $T_c$ for CrSbSe$_3$. (e) Field dependence of the maximum magnetic entropy change $-\Delta S_M^{max}$ and the relative cooling power (RCP) with power-law fitting in solid lines along all three axes for CrSbSe$_3$.}
\label{3}
\end{figure}

To further characterize the anisotropic magnetic properties of CrSbSe$_3$, the isothermal magnetization with field up to 50 kOe applied along each axis from 10 to 100 K are presented in Figs. 2(a)-2(c). At high temperature, the curves have linear field dependence. With decreasing temperature, the curves bend with negative curvatures, indicating dominant FM interaction. Based on the classical thermodynamics and Maxwell's relation, the magnetic entropy change $\Delta S_M(T,H)$ is given by:\cite{Pecharsky,Amaral}
\begin{equation}
\Delta S_M = \int_0^H [\frac{\partial S(T,H)}{\partial H}]_TdH = \int_0^H [\frac{\partial M(T,H)}{\partial T}]_HdH,
\end{equation}
where $[\partial S(T,H)/\partial H]_T$ = $[\partial M(T,H)\partial T]_H$ is based on the Maxwell's relation. For magnetization measured at small temperature and field intervals,
\begin{equation}
\Delta S_M = \frac{\int_0^HM(T_{i+1},H)dH-\int_0^HM(T_i,H)dH}{T_{i+1}-T_i}.
\end{equation}
The calculated $-\Delta S_M(T,H)$ are presented in Figs. 2(d)-2(f). All the curves exhibit a peak feature near $T_c$. Peaks broads asymmetrically on both sides with increasing field. The maximum of $-\Delta S_M(T,H)$ reach 2.16 J kg$^{-1}$ K$^{-1}$, 2.11 J kg$^{-1}$ K$^{-1}$, and 2.03 J kg$^{-1}$ K$^{-1}$ for $a$, $b$, and $c$ axis, respectively. It should be noted that all the values of $-\Delta S_M(T,H)$ for easy $a$ axis are positive, however, the values for hard $b$ and $c$ axes are negative at low temperatures in low fields stemming from a strong temperature-dependent magnetic anisotropy.\cite{Liu02}

\begin{figure}
\centerline{\includegraphics[scale=1]{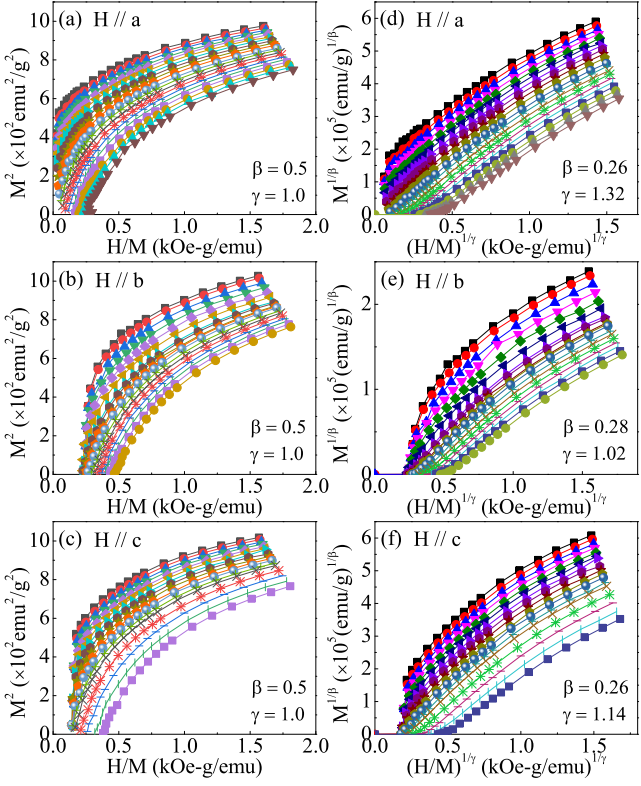}}
\caption{(Color online). The Arrott plot of $M^2$ vs $H/M$ (a-c) and the modified Arrott plot of $M^{1/\beta}$ vs $(H/M)^{1/\gamma}$ (d-f) with indicated $\beta$ and $\gamma$ for $a$, $b$ and $c$ axis, repsectively.}
\label{4}
\end{figure}

Based on a generalized scaling analysis,\cite{Franco} the normalized magnetic entropy change, $\Delta S_M/\Delta S_M^{max}$, estimated for each constant field, is scaled to the reduced temperature $t$ as defined in the following equations:
\begin{equation}
t_- = (T_{peak}-T)/(T_{r1}-T_{peak}), T<T_{peak},
\end{equation}
\begin{equation}
t_+ = (T-T_{peak})/(T_{r2}-T_{peak}), T>T_{peak},
\end{equation}
where $T_{r1}$ and $T_{r2}$ are the lower and upper reference temperatures at half-width full maximum of $\Delta S_M/\Delta S_M^{max}$. As we can see, the normalized $\Delta S_M/\Delta S_M^{max}$ near $T_c$ collapses onto a universal curve at the indicated fields for all three axes [Figs. 3(a)-3(c)], indicating the second-order PM-FM transition in CrSbSe$_3$. The ineligible deviation at low temperatures along the hard $b$ axis is mostly contributed by the magnetocrystalline anisotropy effect. Based on the Stoner-Wolfarth model,\cite{Stoner} the magnetocrystalline anisotropy constant $K_u$ can be estimated from the saturation regime in the isothermal magnetization curves. Within this model the value of $K_u$ in single domain state is related to the saturation magnetization $M_s$ and the saturation field $H_s$ with $\mu_0$ is the vacuum permeability:
\begin{equation}
\frac{2K_u}{M_s} = \mu_0H_{s}.
\end{equation}
When $H\parallel b$, the anisotropy becomes maximal. We estimated the $M_s$ by using a linear fit of $M(H)$ above 20 kOe for $H\parallel b$, which monotonically decreases with increasing temperature [inset in Fig. 3(d)]. Then we determined the $H_s$ as the intersection point of two linear fits: one being a fit to the saturated regime at high field, and the other being a fit of the unsaturated linear regime at low field. The values of $H_s$ share a similar temperature dependence [inset in Fig. 3(d)], resulting in a strongly temperature-dependent $K_u$ [Fig. 3(d)]. The calculated $K_u$ is $\sim$ 148.5 kJ m$^{-3}$ at 10 K for CrSbSe$_3$, which is much larger than that of Cr$_2$(Si,Ge)$_2$Te$_6$,\cite{Liu02} and comparable with that of Cr(Br,I)$_3$.\cite{Richter}

\begin{table}
\caption{\label{tab} The values of magnetic entropy change ($-\Delta S_M^{max}$) and relative cooling power (RCP) at field change of 50 kOe. Critical exponents of CrSbSe$_3$ along $a$, $b$, and $c$ axis, recpectively. The MAP, KFP, and CI represent the modified Arrott plot, the Kouvel-Fisher plot, and the critical isotherm, respectively.}
\begin{ruledtabular}
\begin{tabular}{lllllll}
   & $-\Delta S_M^{max}$ & RCP & $T_c$ & $\beta$ & $\gamma$ & $\delta$ \\
   & (J/kg-K) & (J/kg) &  &  &  &  \\
   \hline
   $H\parallel a$ & 2.16 & 163.1 & & & &\\
   MAP & & & 70.2(1) & 0.26(1) & 1.32(2) & 6.08(12)\\
   KFP & & & 70.2(1) & 0.26(1) & 1.33(2) & 6.12(12)\\
   CI  & & & & & & 6.17(9) \\
   \hline
   $H\parallel b$ & 2.03 & 142.1 & & & &\\
   MAP & & & 70.1(2) & 0.28(2) & 1.02(1) & 4.64(22)\\
   KFP & & & 70.3(2) & 0.32(2) & 1.03(1) & 4.22(17)\\
   CI  & & & & & & 4.14(16)\\
   \hline
   $H\parallel c$ & 2.11 & 154.1 & & & &\\
   MAP & & & 69.8(2) & 0.26(1) & 1.14(5) & 5.38(2)\\
   KFP & & & 69.8(2) & 0.25(1) & 1.19(4) & 5.76(2)\\
   CI  & & & & & & 5.35(17)\\
\end{tabular}
\end{ruledtabular}
\end{table}

\begin{figure}
\centerline{\includegraphics[scale=1]{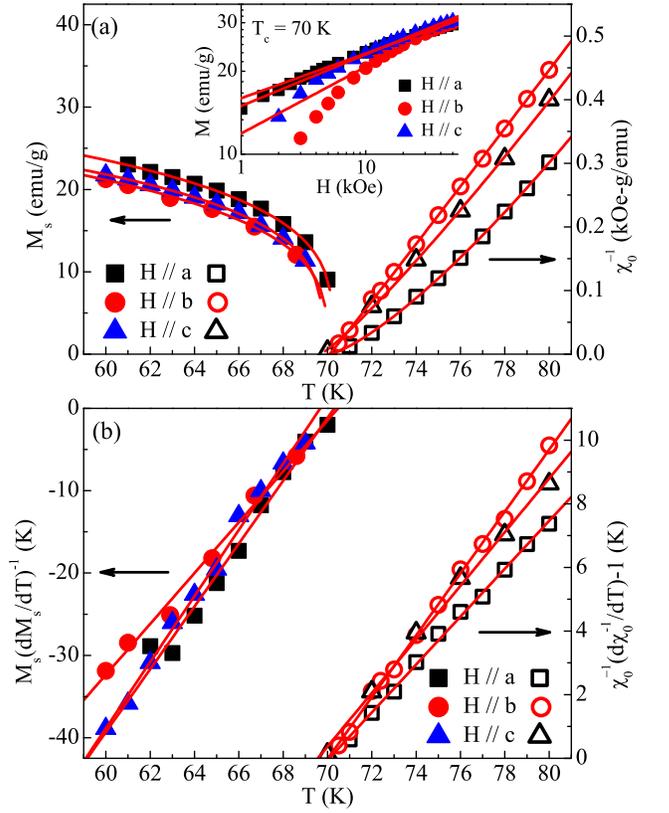}}
\caption{(Color online). (a) Temperature dependence of the spontaneous magnetization $M_s$ (left) and the inverse initial susceptibility $\chi_0^{-1}$ (right) with solid fitting curves for CrSbSe$_3$. Inset shows the log$_{10}M$ vs log$_{10}H$ at 70 K with linear fitting curve. (b) Kouvel-Fisher plots of $M_s(dM_s/dT)^{-1}$ (left) and $\chi_0^{-1}(d\chi_0^{-1}/dT)^{-1}$ (right) with solid fitting curves for CrSbSe$_3$.}
\label{5}
\end{figure}

Another parameter to characterize the potential magnetocaloric effect of materials is the relative cooling power (RCP):\cite{Gschneidner}
\begin{equation}
RCP = -\Delta S_M^{max} \times \delta T_{FWHM},
\end{equation}
where the FWHM denotes the full width at half maximum of $-\Delta S_M$ curve. The $RCP$ corresponds to the amount of heat which could be transferred between cold and hot parts of the refrigerator in an ideal thermodynamic cycle. The calculated RCP values of CrSbSe$_3$ reach maxima at 50 kOe of 163.1 J kg$^{-1}$, 142.1 J kg$^{-1}$, and 154.1 J kg$^{-1}$ for $a$, $b$, and $c$ axis, respectively [Fig. 3(e)]. In addition, the field dependence of $-\Delta S_M^{max}$ and RCP can be well fitted by using the power-law relations $-\Delta S_M^{max} = aH^n$ and $RCP = bH^m$ [Fig. 3(e)].\cite{VFranco}

For a second-order PM-FM phase transition, the spontaneous magnetization ($M_s$) below $T_c$, the initial magnetic susceptibility ($\chi_0^{-1}$) above $T_c$, and the field-dependent magnetization (M) at $T_c$ can be characterized by a set of critical exponents $\beta$, $\gamma$, and $\delta$, respectively.\cite{Stanley} The mathematical definitions of the exponents from magnetization measurement are given below:
\begin{equation}
M_s (T) = M_0(-\varepsilon)^\beta, \varepsilon < 0, T < T_c,
\end{equation}
\begin{equation}
\chi_0^{-1} (T) = (h_0/m_0)\varepsilon^\gamma, \varepsilon > 0, T > T_c,
\end{equation}
\begin{equation}
M = DH^{1/\delta}, T = T_c,
\end{equation}
where $\varepsilon = (T-T_c)/T_c$; $M_0$, $h_0/m_0$ and $D$ are the critical amplitudes.\cite{Fisher} For the original Arrott plot, $\beta$ = 0.5 and $\gamma$ = 1.0.\cite{Arrott1} Based on this, the magnetization isotherms $M^2$ vs $H/M$ should be a set of parallel straight lines. The isotherm at the critical temperature $T_c$ should pass through the origin. As shown in Figs. 4(a-c), all curves in the Arott plot of CrSbSe$_3$ are nonlinear, with a downward curvature, indicating that the Landau mean-field model is not applicable to CrSbSe$_3$. From the Banerjee$^\prime$s criterion,\cite{Banerjee} the first (second) order phase transition corresponds to a negative (positive) slope. Thus, the downward slope confirms its a second-order PM-FM transition in CrSbSe$_3$. In the more general case, the Arrott-Noaks equation of state provides a modified Arrott plot:\cite{Arrott2}
\begin{equation}
(H/M)^{1/\gamma} = a\varepsilon+bM^{1/\beta},
\end{equation}
where $\varepsilon = (T-T_c)/T_c$ and $a$ and $b$ are fitting constants. Figures 4(d-f) present the modified Arrott plots for all three axes with a self-consistent method,\cite{Kellner, Pramanik} showing a set of parallel quasi-straight lines at high field.

\begin{figure}
\centerline{\includegraphics[scale=1]{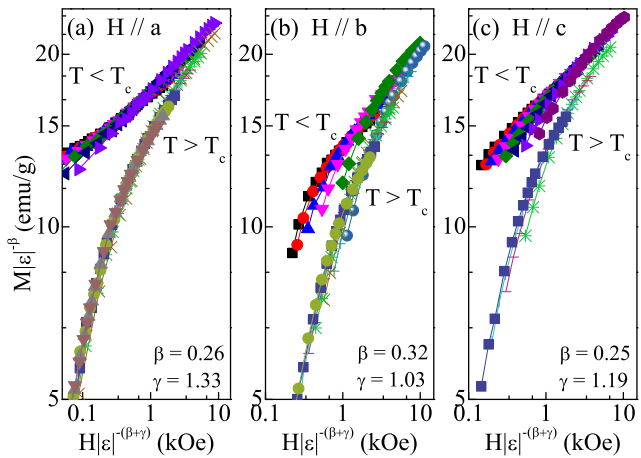}}
\caption{(Color online). Scaling plots of renormalized $m = M|\varepsilon|^{-\beta}$ vs $h = H|\varepsilon|^{-(\beta+\gamma)}$ above and below $T_c$ for CrSbSe$_3$.}
\label{6}
\end{figure}

To obtain the anisotropic critical exponents $\beta$, $\gamma$ and $\delta$, the linearly extrapolated $M_s(T)$ and $\chi_0^{-1}(T)$ against temperature are plotted in Fig. 5(a). According to Eqs. (7) and (8), the solid fitting lines give that $\beta = 0.26(1)$ and $\gamma = 1.32(2)$ for easy $a$ axis, close to the reported values ($\beta = 0.25$ and $\gamma = 1.38$).\cite{Kong} For the $b$ axis, $\beta = 0.28(2)$ and $\gamma = 1.02(1)$, while for the $c$ axis, $\beta = 0.26(1)$ and $\gamma = 1.14(5)$. This lies between the values of theoretical tricritical mean field model ($\beta$ = 0.25 and $\gamma$ = 1.0) and 3D Ising model ($\beta$ = 0.325 and $\gamma$ = 1.24).\cite{Khuang,LeGuillou} The value of $\beta$ is outside of the window $0.1 \leq \beta \leq 0.25$ for a 2D magnet,\cite{Taroni} suggesting a 3D magnetic behavior for quasi-1D CrSbSe$_3$. According to Eq. (9), the $M(H)$ at $T_c$ should be a straight line in log-log scale with the slope of $1/\delta$. Such fitting yields $\delta$ = 6.17(9), 4.14(16), and 5.35(17), for $a$, $b$, and $c$ axis, respectively, which agrees well with the calculated values from $\beta$ and $\gamma$ based on the Widom relation $\delta = 1+\gamma/\beta$.\cite{Widom} The self-consistency can also be checked via the Kouvel-Fisher method,\cite{Kouvel} where $M_s(T)/(dM_s(T)/dT)^{-1}$ and $\chi_0^{-1}(T)/(d\chi_0^{-1}(T)/dT)^{-1}$ plotted against temperature should be straight lines with slopes $1/\beta$ and $1/\gamma$, respectively. The linear fits to the plots [Fig. 5(b)] yield $\beta = 0.26(1)$ and $\gamma = 1.33(2)$ for $a$ axis, $\beta = 0.32(2)$ and $\gamma = 1.03(1)$ for $b$ axis, $\beta = 0.25(1)$ and $\gamma = 1.19(4)$ for $c$ axis, respectively, very close to the values obtained from modified Arrott plot. All the critical exponents obtained from different methods are listed in Table I. It seems that the critical behavior of CrSbSe$_3$ is much different along the different axis and cannot be described by any single model. However, it is clear that 3D critical behavior dominates in quasi-1D CrSbSe$_3$ and the strong magnetocrystalline anisotropy in CrSbSe$_3$ plays an important role in the origin of anisotropic critical exponents.

Scaling analysis can be used to estimate the reliability of the obtained critical exponents. According to scaling hypothesis, the magnetic equation of state in the critical region obeys a scaling relation can be expressed as:
\begin{equation}
M(H,\varepsilon) = \varepsilon^\beta f_\pm(H/\varepsilon^{\beta+\gamma}),
\end{equation}
where $f_+$ for $T > T_c$ and $f_-$ for $T < T_c$, respectively, are the regular functions. In terms of the variable $m\equiv\varepsilon^{-\beta}M(H,\varepsilon)$ and $h\equiv\varepsilon^{-(\beta+\gamma)}H$, renormalized magnetization and renormalized field, respectively, Eq.(10) reduces to a simple form:
\begin{equation}
m = f_\pm(h).
\end{equation}
It implies that for a true scaling relation with the proper selection of $\beta$, $\gamma$, and $\delta$, the renormalized $m$ versus $h$ data will fall onto two different universal curves; $f_+(h)$ for temperature above $T_c$ and $f_-(h)$ for temperature below $T_c$. Using the values of $\beta$ and $\gamma$ obtained from the Kouvel-Fisher plot, we have constructed the renormalized $m$ vs $h$ plots in Fig. 6. It is clear seen that all the experimental data collapse onto two different branches: one above $T_c$ and another below $T_c$, confirming proper treatment of the critical regime.

\section{CONCLUSIONS}

In summary, we have studied in details the anisotropic magnetocaloric effect and critical behavior of CrSbSe$_3$ single crystal. The second-order in nature of the PM-FM transition near $T_c$ = 70 K has been verified by the scaling analysis of magnetic entropy change $\Delta S_M$. A large magnetocrystalline anisotropy constant $K_u$ is estimated to be 148.5 kJ m$^{-3}$ at 10 K, comparable with that of Cr(Br,I)$_3$. A set of critical exponents $\beta$, $\gamma$, and $\delta$ along each axis estimated from various techniques match reasonably well and follow the scaling equation, indicating a 3D magnetic behavior in CrSbSe$_3$. Further neutron scattering and theoretical studies are needed to shed more light on the anisotropic magnetic coupling in low dimensions.

\section*{Acknowledgements}
This work was supported by the US DOE-BES, Division of Materials Science and Engineering, under Contract No. DE-SC0012704 (BNL).

\end{document}